# Exploring the Effects of Entanglement on Quantum Machine Learning of Pathogen Epitope-Receptor Binding

Aspen Erlandsson Brisebois[1,2], Luis Pablo Gonzalez Dominguez[1,3,4], Shivansi Prajapati[4,5], Zahed Khatooni[1], Heather L. Wilson[1], Connor Burbridge[6], Brook Byrns[6], Sureesh Tikoo[1,7], Christophe Pere[8], Steven Rayan[3,4*], Gordon Broderick[1,3,4*]

[1] Vaccine and Infectious Disease Organization (VIDO), University of Saskatchewan, Saskatoon, SK, Canada

[2] Engineering Science, Faculty of Applied Science, University of Toronto, Toronto, ON, Canada

[3] Department of Mathematics and Statistics, University of Saskatchewan, Saskatoon, SK, Canada

[4] Centre for Quantum Topology and Its Applications (quanTA), University of Saskatchewan, Saskatoon, SK, Canada

[5] Department of Computer Science, University of Saskatchewan, Saskatoon, SK, Canada

[6] Advanced Research Computing, University of Saskatchewan, Saskatoon, SK, Canada

[7] Vaccinology and Immunotherapeutics Program, School of Public Health, University of Saskatchewan, Saskatoon, SK, Canada

[8] Nord quantique, Sherbrooke, QC, Canada

* Joint corresponding authors: steven.rayan@usask.ca ; gordon.broderick@usask.ca

**Abstract**

Parameterized quantum circuits (PQCs) provide a flexible substrate for hybrid quantum machine learning (QML), but their practical value on Noisy Intermediate-Scale Quantum (NISQ) devices remains an empirical question, especially because training depth and scale can introduce optimization challenges such as barren plateaus. Here we study how the number and topology of two-qubit entangling gates in the feature-map stage influence a fixed hybrid QNN workflow for classifying strong versus weak epitope-receptor binding in Porcine Reproductive and Respiratory Syndrome (PRRS) vaccine design. The dataset consists of docking-derived binding affinities for N=80 9-mer epitopes, labeled as Strong or Weak binding, and partitioned into training, validation, and test subsets using a 40:30:30 split. We compare a classical CNN benchmark with a hybrid Embedding-QNN architecture under four feature-map configurations: a non-entangling Z feature map, an all-to-all high-entanglement ZZ feature map, and two interleaved nearest-neighbor entanglement patterns of low and high depth. Using classification accuracy as the primary endpoint for this nearly balanced binary dataset, together with normalized area under the accuracy curve (AUAC) and train-test comparisons, we find that unfiltered QNN runs perform near chance (~53-54% test accuracy). In contrast, validation-gated run filtering at an 85% validation-accuracy threshold identifies QNN runs whose test-set performance is comparable to the classical CNN benchmark (~70-75%). Among the configurations tested, the high-entanglement ZZ feature map provides the strongest evidence of reduced training-set overfit, with a lower training AUAC and the highest test/training AUAC ratio, while preserving competitive test-set accuracy. These results do not establish a general QML advantage, but they suggest that feature-map entanglement topology is a meaningful design variable for sparse biological screening tasks and warrants further evaluation with additional metrics, larger datasets, and noise-aware or hardware-based experiments.

# I. Introduction

Quantum computing is moving from a purely theoretical discipline toward experimentally accessible, application-oriented platforms. In the NISQ era, gate-based devices contain enough qubits to support nontrivial experiments, but their circuit depth, gate fidelity, and scalability remain limited [Preskill, 2018]. Parameterized quantum circuits are therefore attractive for hybrid QML because they can be trained with classical optimization loops on circuits that are relatively shallow, but this same training paradigm is also vulnerable to flat or exponentially suppressed gradient landscapes in sufficiently random or deep circuits [McClean et al., 2018]. Accordingly, claims of practical QML utility should be framed cautiously: the relevant question is not whether QML is generically superior to classical learning, but whether a specific quantum circuit family, feature map, optimizer, and dataset together produce useful predictive structure under realistic resource constraints. Recent work on QML generalization from limited training examples provides theoretical and numerical reasons to examine this question in small-data regimes, while also emphasizing that any favorable behavior depends on the model family and the data distribution [Caro et al., 2022; Gil-Fuster et al., 2024]. This small-data setting is especially relevant to molecular and immunological screening, where high-fidelity simulations and experimental validation can be slow, costly, and operationally constrained. For example, molecular-scale simulations of viral structures can require very large high-performance computing allocations [Pezeshkian et al., 2023]. As a result, related machine-learning approaches are increasingly being explored for drug discovery, molecular docking, and protein-ligand binding prediction [von Lilienfeld et al., 2020; Haque et al., 2025; Shu et al., 2024; Jeong et al., 2025].

In this work, we investigate whether the degree and topology of entanglement in the feature-map stage affect the performance of a fixed hybrid QNN architecture for sparse epitope-receptor binding classification. The available labels are derived from medium-resolution in silico docking experiments, each requiring substantial classical compute time, so the study is intentionally framed as a limited-data screening experiment rather than as a claim of broad QML advantage. The workflow uses a classical learned embedding approach to reduce the one-hot epitope representation and maps the reduced features into a PQC through one of four feature-map configurations. A fixed quantum convolutional/pooling and RealAmplitudes ansatz stage are the applied followed by a classical output layer to assign each epitope to a Strong or Weak binding class. The central comparison is therefore not among distinct QML model classes such as QNNs versus quantum kernels, but among feature-map entanglement configurations within the same Embedding-QNN family which are then benchmarked against a simple CNN classifier.

The paper is organized as follows. Section II (Methods) describes the dataset, the classical CNN benchmark, the fixed Embedding-QNN architecture, the four feature-map entanglement configurations, and the training and validation-gating procedure. Section III (Results) reports unfiltered and validation-filtered performance, including AUAC summaries and graph-theoretic descriptions of the tested entanglement patterns. In Section IV (Discussion) we interpret the results and discuss the limitations of the work, and in Section V (Conclusion) we summarize the implications for sparse vaccine-antigen screening and future QML experiments.

# II. Methods

## *A. Dataset Preparation*

The dataset consists of binding affinities predicted using extensive molecular docking analysis between a commonly expressed swine leukocyte antigen (SLA) histocompatibility receptor and N=80 epitopes, each 9 amino acids in length, extracted from10 proteins whose sequences are conserved across multiple Porcine Reproductive and Respiratory Syndrome Virus (PRRSV) strains.

SLA class I (SLA-1) receptor molecules play a dominant role in the presentation of cytotoxic T lymphocyte (CTL) immune epitopes, and the allele studied here is broadly expressed across five swine breeds [Zhang et al., 2011; Lunney et al., 2009; Ho et al., 2009]. The epitopes consisted of 9-amino-acid sequences selected from 10 proteins expressed by multiple PRRS virus strains. For each 9-mer primary sequence, multiple corresponding three-dimensional (3D) conformations were computed using AlphaFold 3 [Abramson et al., 2024], and the conformation judged most suitable was selected for subsequent docking experiments [Jiang et al., 2025; Alotaiq et al., 2025]. Docking was performed using the HADDOCK 2.4 platform [Honorato et al., 2024]. Applying Otsu's thresholding algorithm [Otsu, 1979], binding-affinity values were assigned to Strong (<= -77.8; n=38) or Weak (> -77.8; n=42) binding-affinity classes. A one-hot encoding scheme was used to convert each 9-mer sequence into a two-dimensional (2D) numerical array, resulting in a 20 x 9 input feature space. For example, the epitope RVPILRTVF, with a binding energy of -92.5 against the SLA-1 receptor and therefore labeled Strong, is represented as the encoded array in Figure 1.

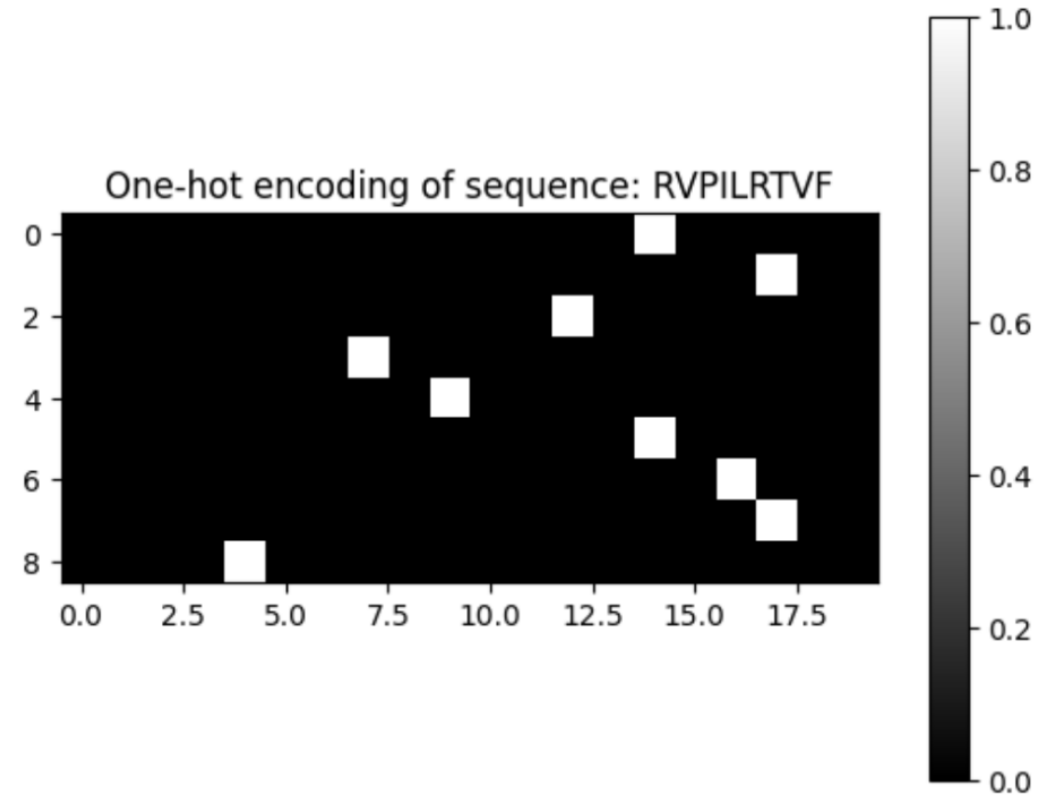

**Figure 1:** One-hot encoding scheme applied to the epitope *RVPILRTVF*

Because in vivo measurements and high-fidelity computational screens are resource-intensive, biological datasets of this kind are often sparse and limited in scope. To test performance under a deliberately conservative data regime, we partitioned the 80 example epitopes into training, validation, and test subsets using a 40:30:30 split [Kjeldsberg et al., 2025]. The full dataset is nearly balanced rather than strongly imbalanced, with 38 Strong and 42 Weak labels. Consequently, all randomized epitope shuffling and dataset splitting were performed in a stratified manner so that each subset retained approximately this same balanced class representation.

### *B. Classification Model Construction*

We compare one classical benchmark architecture with one hybrid QNN architecture evaluated under four feature-map configurations. The benchmark is a classical convolutional neural network (CNN) classifier [Chen et al., 2025]. The QML component is not a family of different QML model classes; rather, it is a fixed Embedding-QNN workflow in which only the feature-map entanglement pattern and depth are varied. This distinction is important because the core study is the effect of entanglement in the feature map, not a comparison of QNNs with quantum kernels or other QML paradigms. The general structure of the benchmark CNN was kept simple, with no node dropout, no feature pooling [Zabin et al., 2025], and dense linear layers for classification as shown in Figure 2. Various degrees of node dropout were tested and did not noticeably affect classification performance for this dataset, likely because of the limited scale of the CNN model.

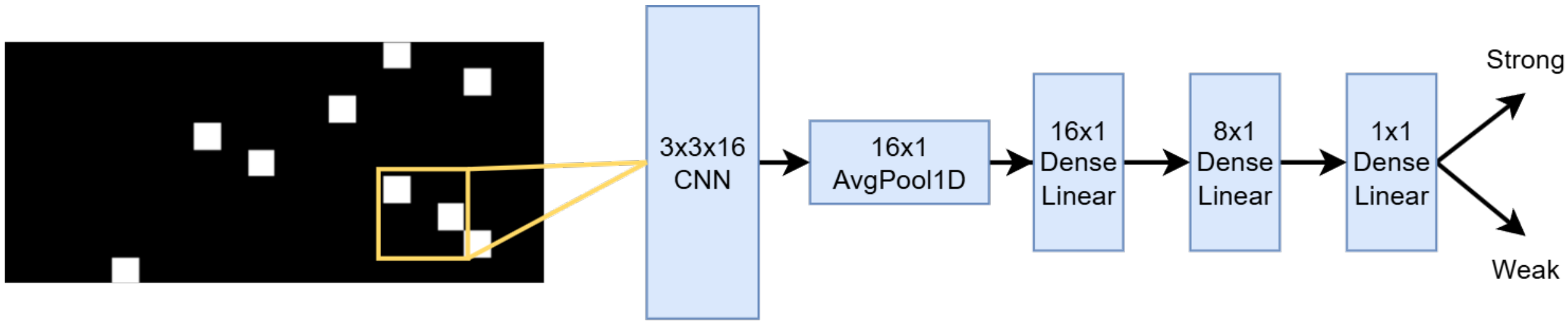

**Figure 2:** CNN Classifier Benchmark Model

The hybrid QML architecture (Fig. 3) consists of an initial classical feature-embedding stage for dimensional reduction, followed by a parameterized QNN circuit comprising a feature map, a quantum convolutional stage, a quantum pooling stage, a variational ansatz stage, qubit measurement, and a final classical output layer. The design is therefore hybrid throughout: the embedding and output weights are trained classically, while the central feature transformation is implemented by a PQC whose feature-map entanglement is the experimental variable of interest.

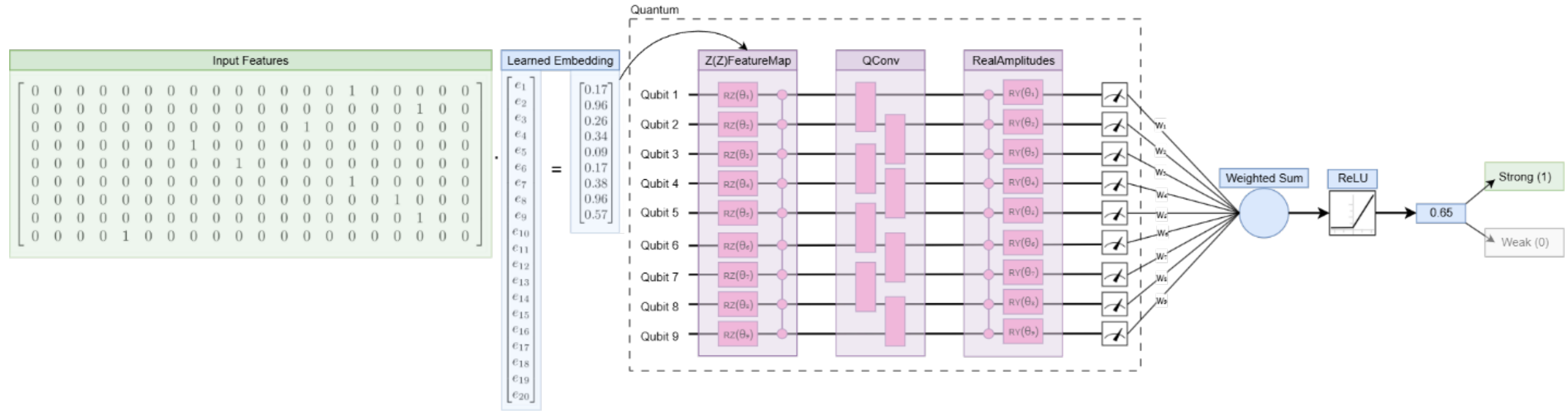

**Figure 3:** QML High-Level Architecture

Since the one-hot encoding of the 9-mer epitopes produces an input space of 180 elements (9 x 20), we first perform dimensional reduction using a sparse autoencoder to generate a 20 x 1 learned embedding vector [Gallifant et al., 2025]. Though not necessarily optimal this dimensional reduction provides a compact learned representation suitable for loading epitope characteristics into a small, simulated circuit while preserving an interpretable amino-acid-level encoding. The reduced vector is then translated onto the quantum circuit using a feature map [Singh and Pokhrel, 2025] (Fig. 4). In the Z feature map, input features are encoded through single-qubit phase rotations and no additional two-qubit feature-map entangling gates are introduced. In the ZZ feature map, pairs of input features enter through two-qubit phase interactions, producing an entangling feature space of the type introduced in quantum-enhanced feature-space classification [Havlíček et al., 2019]. In this implementation, we use the ZFeatureMap and ZZFeatureMap classes available in Qiskit as convenient software realizations of these feature-map forms; the mathematical distinction is the feature-map structure itself, not an IBM-specific or Qiskit-specific claim. We compare four feature-map configurations: a non-entangling Z baseline, a high-entanglement all-to-all ZZ feature map with two repetitions, and two Z-based interleaved nearest-neighbor entanglement patterns with low and high two-qubit-gate counts [Anand, 2022]. The interleaved patterns serve as topology controls: the high-depth interleaved circuit is chosen to have a total number of two-qubit gates comparable to the high-entanglement ZZ circuit, allowing us to distinguish entanglement density from entanglement topology (Fig. 5). The basic features of the four parameterized quantum circuit configurations are summarized in Table 1. These configurations apply only to two-qubit gates at the feature-mapping stage; an additional 40 two-qubit gates arise from the fixed quantum convolutional, pooling, and RealAmplitudes ansatz stages (Fig. 6).

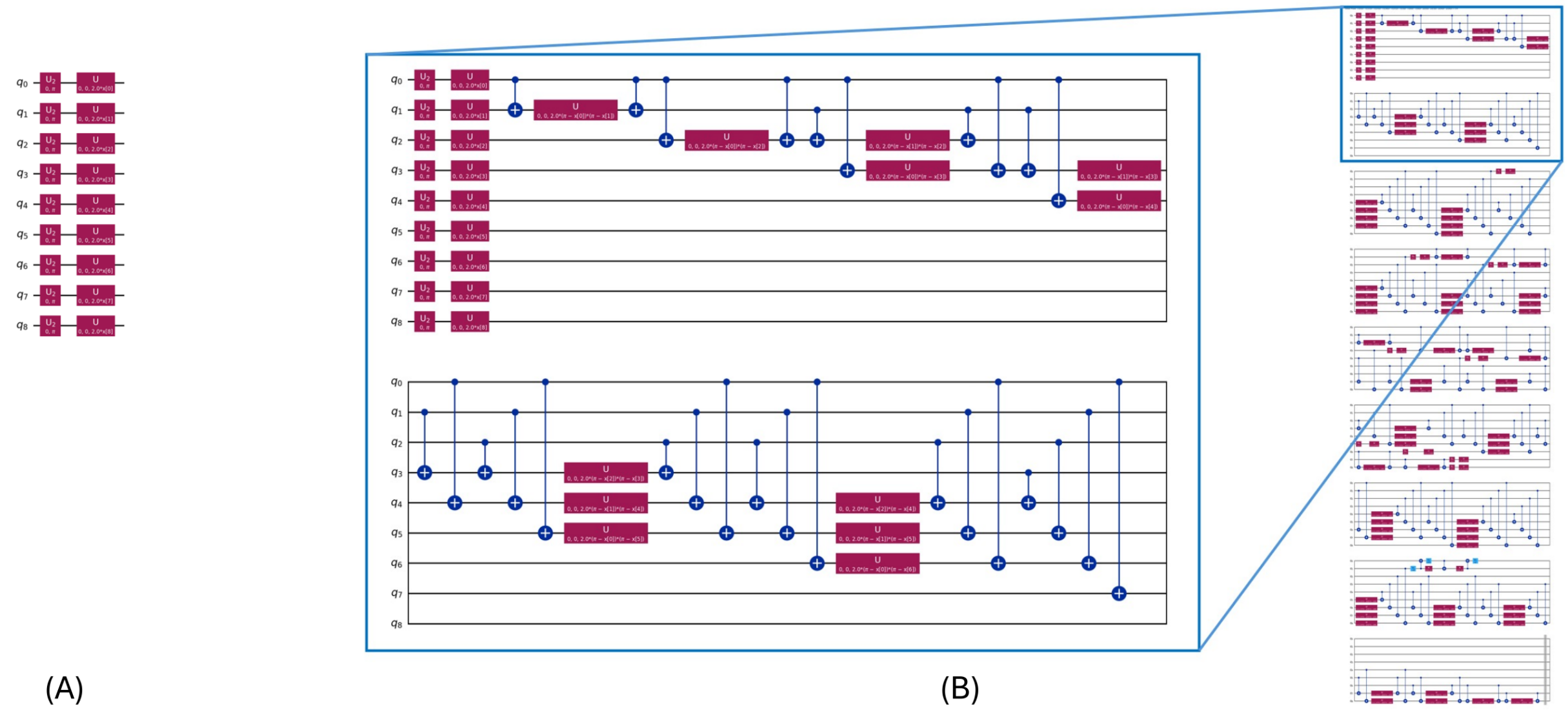

**Figure 4**: Z feature map circuit without feature-map entanglement (A) and high-entanglement ZZ feature map circuit (B).

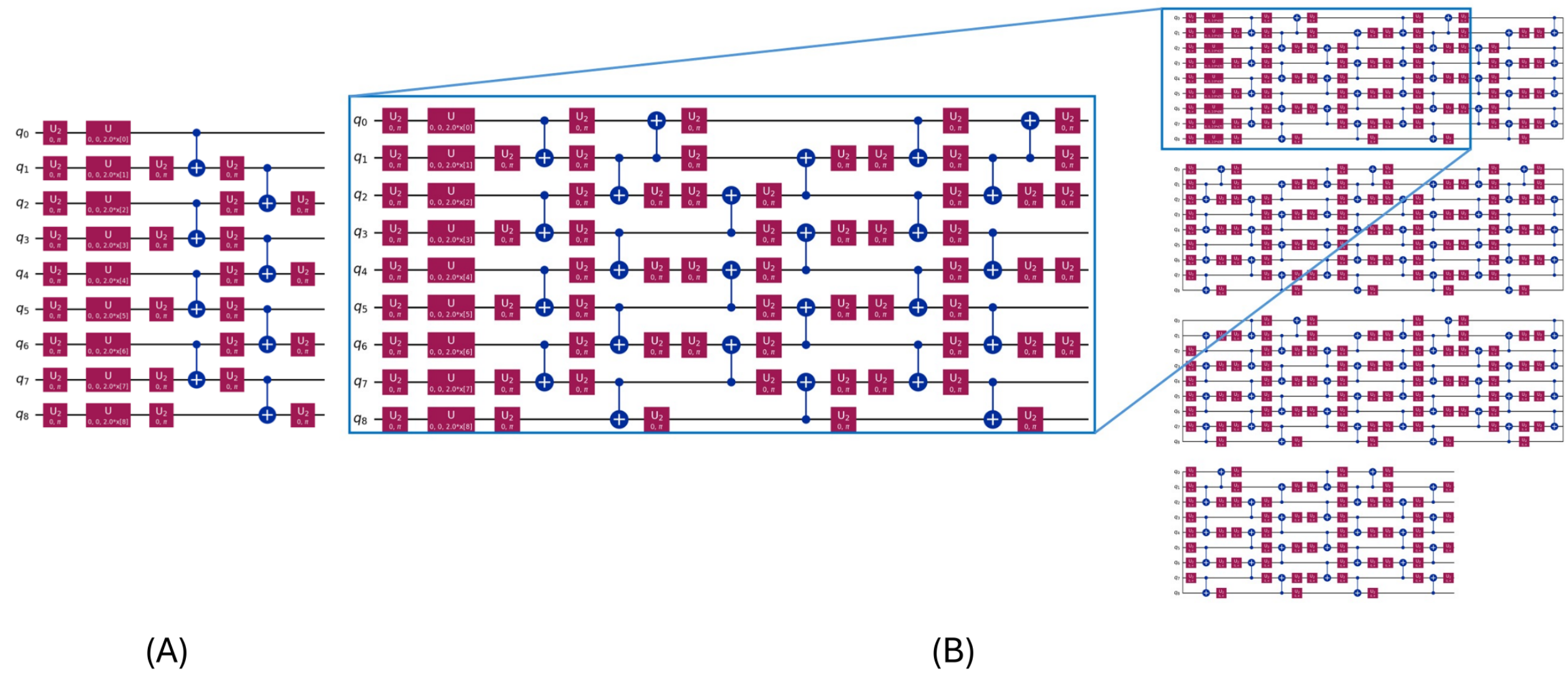

**Figure 5**: Z feature map with low-depth interleaved nearest-neighbor entanglement (A) and Z feature map with high-depth interleaved nearest-neighbor entanglement (B).

This initial feature mapping is followed by a quantum convolutional (QConv) stage and a quantum pooling (QPool) stage (Fig. 6(A) and 6(B)) designed according to a standard topology in which parameterized two-qubit gates are applied in parallel across qubit pairs.

**Table 1**. *Summary of entanglement strategies applied to the feature mapping stage**.

| Feature-map configuration | Feature-map two-qubit gates | Entanglement topology |
|---|---|---|
| Z feature map baseline (1 rep; no feature-map entanglement) | 0 | None |
| ZZ feature map high entanglement (2 reps) | 144 | All-to-all pairwise ZZ feature interactions |
| Z feature map + low-depth interleaved entanglement (1 pass) | 8 | Nearest-neighbor interleaved |
| Z feature map + high-depth interleaved entanglement (9 passes) | 144 | Repeated nearest-neighbor interleaved |

* Note that the QConv + pooling + RealAmplitudes ansatz stage was kept constant across all tests and itself contained 40 additional two-qubit gates.

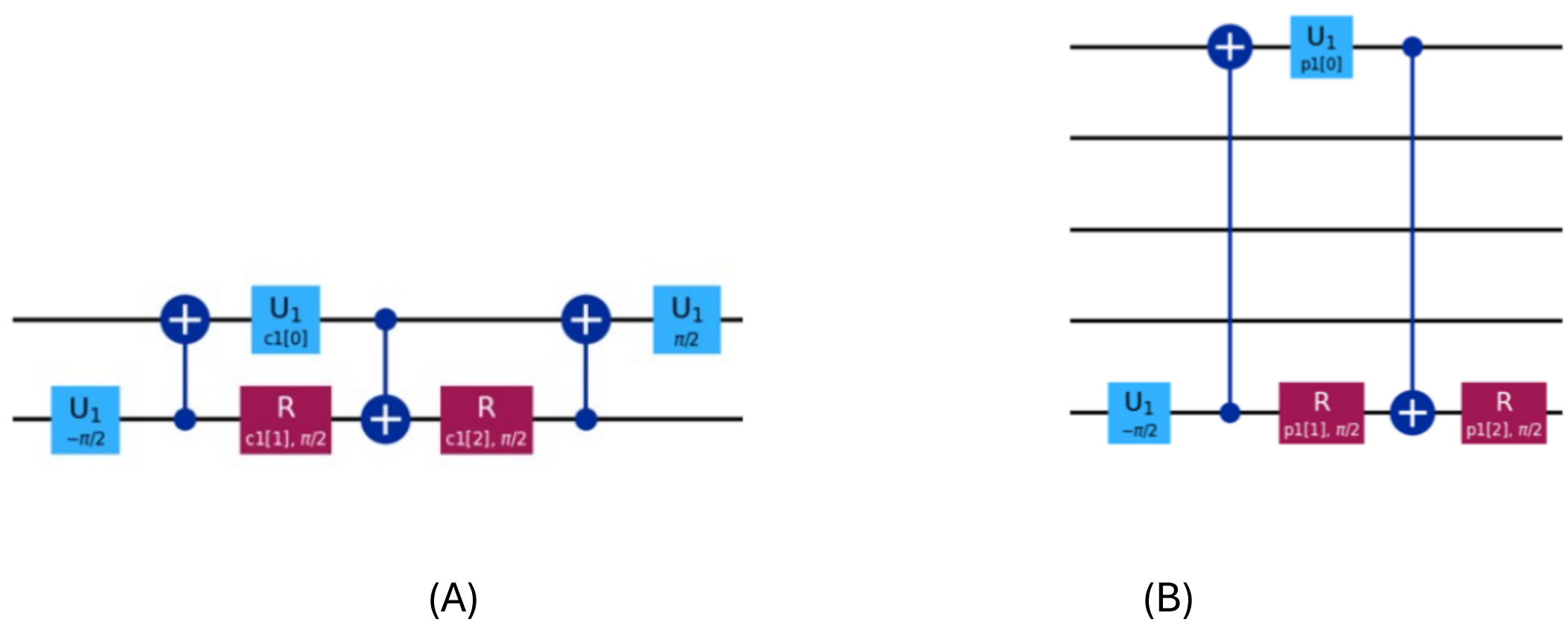

**Figure 6**: Quantum convolutional block (QConv) (A) and quantum pooling block (QPool) (B).

Pooled features then serve as input to the variational ansatz stage, where a Qiskit RealAmplitudes block is used. RealAmplitudes is an ansatz, not itself a classifier; classification is performed only after the QNN measurements are passed to the classical output layer. After the ansatz stage, the circuit is measured to produce 9 floating-point expectation values. Classically learned weights are applied to these values, and a Rectified Linear Unit (ReLU) activation is applied to the weighted sum to assign a Strong or Weak binding class. A complete high-level diagram of the Embedding-QNN structure, including the feature-embedding stage, is shown in Figure 3. While the feature-map configurations vary, the quantum convolutional, pooling, RealAmplitudes ansatz, and classical output components are held constant across all four QNN experiments. The PyTorch [Imambi et al., 2021] Python library was used for classical parameter training and evaluation, with the Qiskit Machine Learning [Sahin et al., 2025] TorchConnector class providing the interface between Qiskit neural-network objects and the PyTorch workflow.

### *C. Training and Evaluation*

All parameter-training experiments used the Adaptive Moment Estimation (Adam) optimizer [Kingma and Ba, 2015], which provided highly consistent behavior in preliminary tests. Unlike gradient-free optimizers often used in QML training, such as COBYLA [Powell, 1964], Adam requires gradients through the hybrid computation graph. In our implementation, quantum-circuit derivatives were obtained by parameter-shift estimation, after which the remaining gradient propagation was handled classically through the PyTorch/Qiskit interface. Simultaneous Perturbation Stochastic Approximation (SPSA) [Li et al., 2023] was also considered, but produced solutions that were too variable to support stable training in this small dataset. Parameter shift was therefore used for all QNN parameter-estimation experiments, an acceptable choice here because the number of trainable quantum parameters is relatively small. Alternative optimizers, including SGD and RMSProp, remain useful candidates for future sensitivity analyses. A constant learning rate of 0.03 was used as a balance between training speed and robust convergence. Each independent training run used a randomized initialization of trainable parameters and a repeated stratified random split of the dataset into training, validation, and test subsets. The corresponding quantum circuit behavior was simulated using the Qiskit StatevectorEstimator.

The simulations conducted in this work do not incorporate stochastic hardware noise, decoherence, readout error, or an explicit mathematical noise model. Instead, we focus on variability arising from uniform-random initialization of PQC parameters and random stratified subsampling into training, validation, and test subsets. All other factors in initialization, training, validation, and testing are deterministic. For each numerical experiment, we train 2,000 independent runs with identical architecture and runtime parameters but different randomized dataset splits and initial parameter values. From this family of 2,000 runs, we retain the subset whose validation-set accuracy exceeds 85% at a given epoch. We refer to this procedure as validation-gated run filtering rather than model selection: it does not compare different QML model classes, and it does not tune the feature map after observing test performance. The threshold is empirical and was chosen because it typically retained approximately 10-30 runs at a given epoch. Because the binary classes are nearly balanced, classification accuracy is a reasonable primary metric for this exploratory study; nevertheless, it is not sufficient on its own to establish QML utility. We therefore also report normalized AUAC summaries and training-versus-test AUAC ratios, and future work should include ROC-AUC, precision-recall behavior, Matthews correlation, calibration, and uncertainty estimates.

## III. Results

The QNN training landscape induced by the chosen circuits and classical gradient-estimation pipeline provides regularization-like behavior in some runs, but it also makes training more fragile than the classical CNN benchmark. Without validation-gated filtering, the average validation and test accuracies of the QNN runs remain close to chance, hovering around 53-54%. After applying the 85% validation-accuracy threshold, the retained QNN runs show much higher test-set performance. The same validation filter was also applied to the classical CNN benchmark runs, although we did not observe a significant positive or negative change in the classical case. For each of the four feature-map entanglement configurations described in Table 1, the average training and test performance of the retained runs is shown with error bars representing one standard error of the mean (SEM).

### *A.* Unfiltered accuracy

When no validation-gated run filtering is applied, the average performance of the QNN configurations explored in this work is low, with test-set accuracy remaining around 53-54% throughout training. This is substantially below the classical CNN benchmark, whose typical test-set accuracy is approximately 70-75% (Figure 7(A)-(D)).

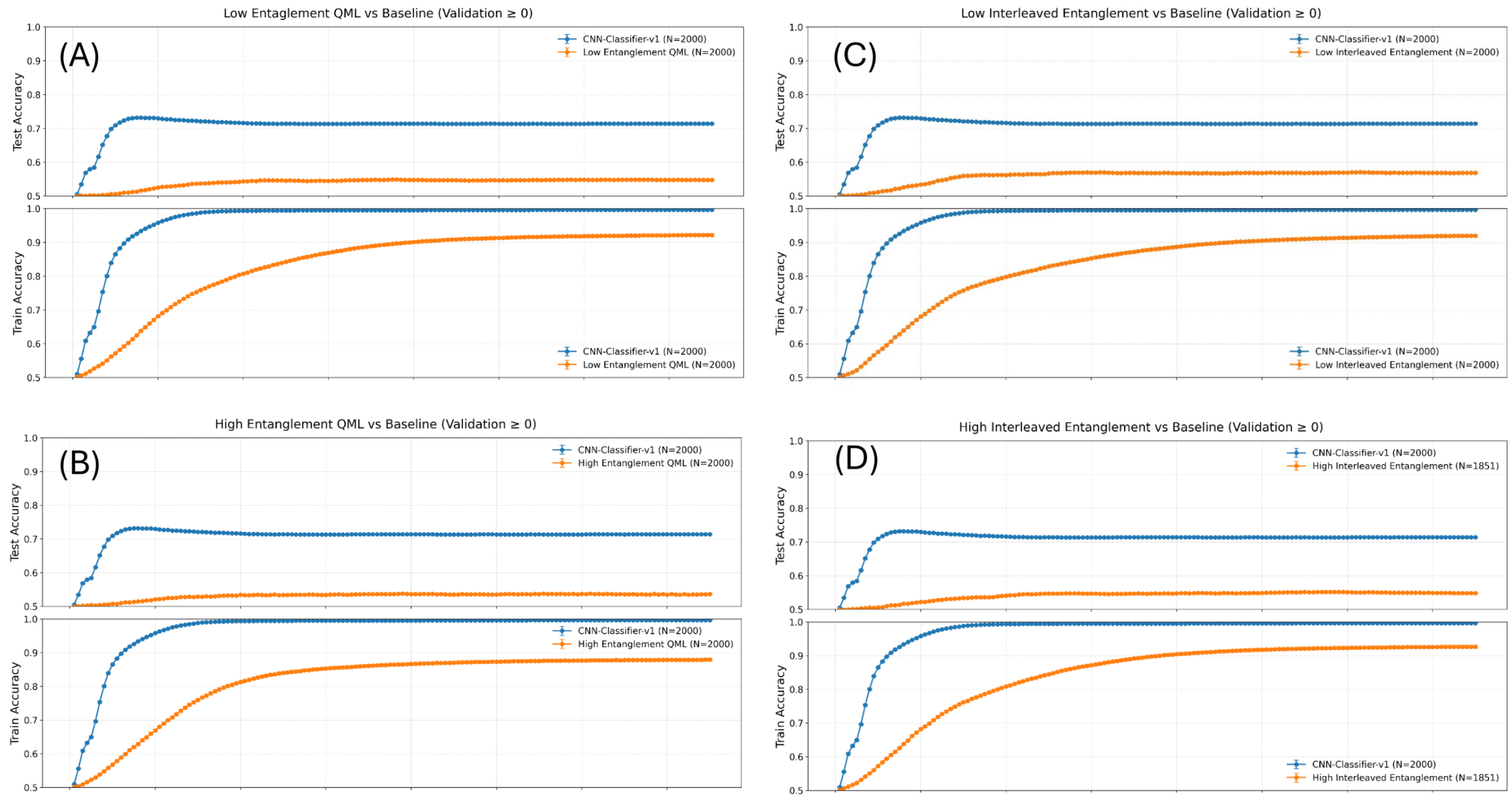

**Figure 7**: Training and test-set accuracy for the classical CNN benchmark (blue) versus the Embedding-QNN architecture (orange) under four feature-map configurations: Z feature map without feature-map entanglement (A), high-entanglement all-to-all ZZ feature map (B), Z feature map with low-depth interleaved entanglement (C), and Z feature map with high-depth interleaved entanglement (D).

**Table 2**. Summary of performance as normalized area under the accuracy curve (AUAC) for both training and test sets. Values are point estimates computed from the mean learning trajectories; epoch-wise SEM is shown in Figure 8 where available.

| Feature-map / benchmark configuration | **AUAC Test Set** | **AUAC Training Set** | **AUAC Test/Training Ratio** |
|---|---|---|---|
| Classical CNN benchmark | 0.704 | 0.983 | 0.716 |
| Embedding-QNN + Z feature map baseline (1 rep; no feature-map entanglement) | 0.699 | 0.942 | 0.708 |
| Embedding-QNN + ZZ feature map high entanglement (2 reps; all-to-all) | 0.712 | 0.899 | 0.792 |
| Embedding-QNN + Z feature map + low-depth interleaved entanglement (1 pass) | 0.703 | 0.932 | 0.754 |
| Embedding-QNN + Z feature map + high-depth interleaved entanglement (9 passes) | 0.707 | 0.940 | 0.751 |

### *B.* Validation-gated run filtering at an 85% validation-accuracy threshold

Applying validation-gated run filtering based on predicted validation-set accuracy produces substantially better test-set performance (Figure 8(A)-(D)). Because only a small fraction of runs survives the filtering stage (~10-30 of 2,000 at a typical epoch), the mean accuracy values from retained-runs oscillate noticeably, with test-set accuracies sometimes above and sometimes below the classical CNN benchmark by more than 5 percentage points. In all cases, 2,000 independent runs were first trained and then filtered according to the validation threshold at the corresponding epoch. A smaller concurrent set of numerical experiments using 1,000 independent runs produced increased variability but similar qualitative trends.

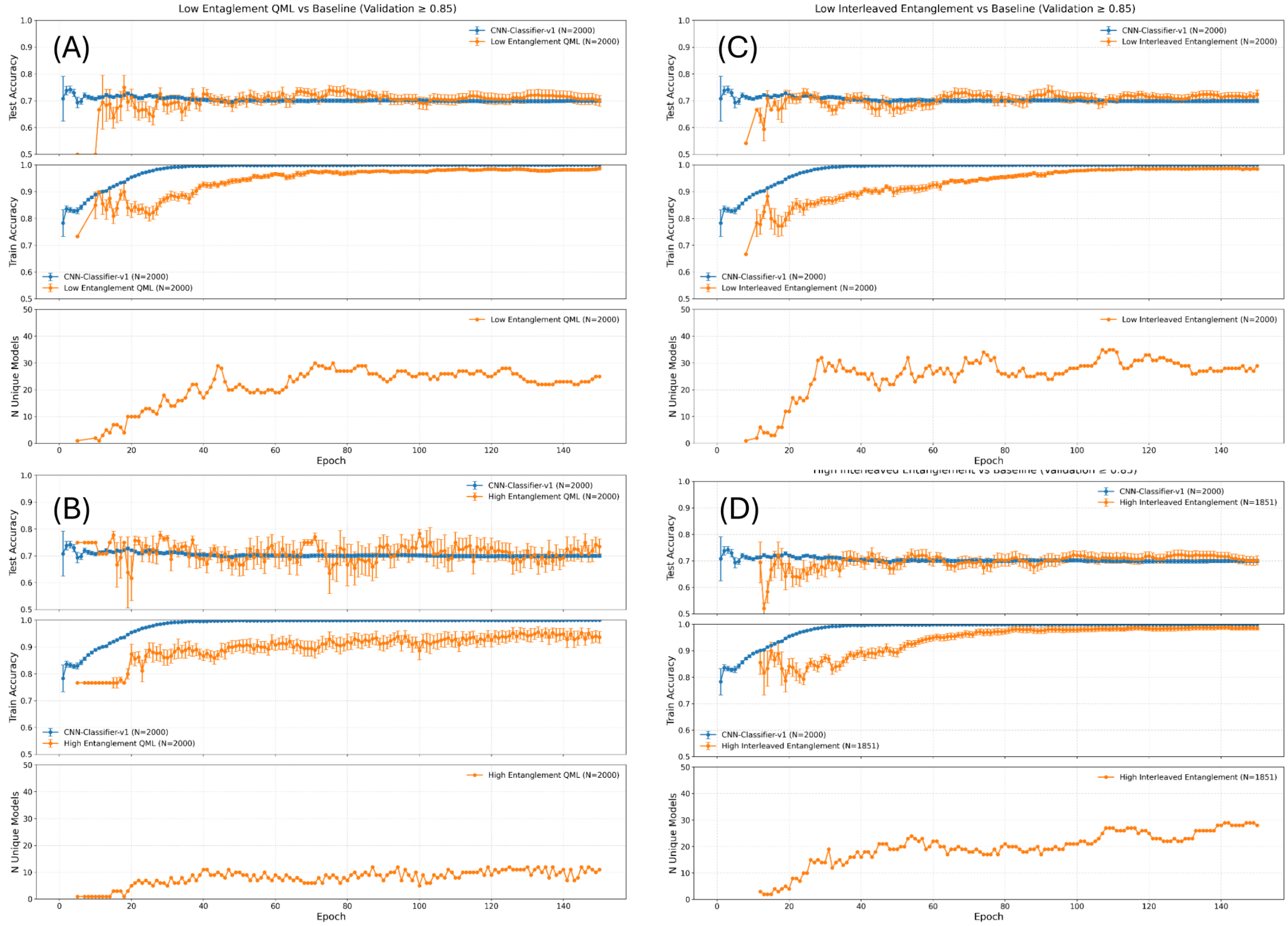

**Figure 8:** Training and test-set accuracy for the classical CNN benchmark (blue) versus the Embedding-QNN architecture (orange) under four feature-map configurations after retaining only runs with validation-set accuracy above 85%: Z feature map without feature-map entanglement (A), high-entanglement all-to-all ZZ feature map (B), Z feature map with low-depth interleaved entanglement (C), and Z feature map with high-depth interleaved entanglement (D).

The learning trajectories shown in Figure 8 suggest that the high-entanglement all-to-all ZZ feature map produces a more conservative fit to the training data among the configurations tested. After an initial rapid increase to approximately 80-85% training accuracy, further improvements occur gradually and eventually plateau near 90% over 150 epochs. In contrast, the other QNN feature-map configurations approach nearly 100% training-set accuracy by epochs 80-100, while the classical CNN reaches near-complete training-set fit within approximately 40 epochs. The high-entanglement ZZ configuration also produces the smallest retained-run subset after validation filtering, with roughly 10 of 2,000 initial runs remaining compared with approximately 20-35 retained runs in the other QNN configurations.

As an aggregate summary across the training horizon, we computed the normalized AUAC for both training and test-set accuracy (Table 2). The four QNN feature-map configurations produce test-set AUAC values (0.699-0.712) similar to the classical CNN benchmark (0.704). However, the corresponding training-set AUAC values are lower for the QNN configurations (0.899-0.942) than for the CNN (0.983). This contrast is captured by the test/training AUAC ratio: most QNN configurations (0.751-0.792) exceed the CNN ratio (0.716), with the exception of the non-entangling Z baseline (0.708). For the high-entanglement ZZ feature map, the combination of the highest test/training AUAC ratio and a lower training-set AUAC suggests that its all-to-all feature-map topology may suppress overfitting more effectively in this dataset. This result should be interpreted as evidence from this specific small-data experiment, not as a general guarantee of QML generalization.

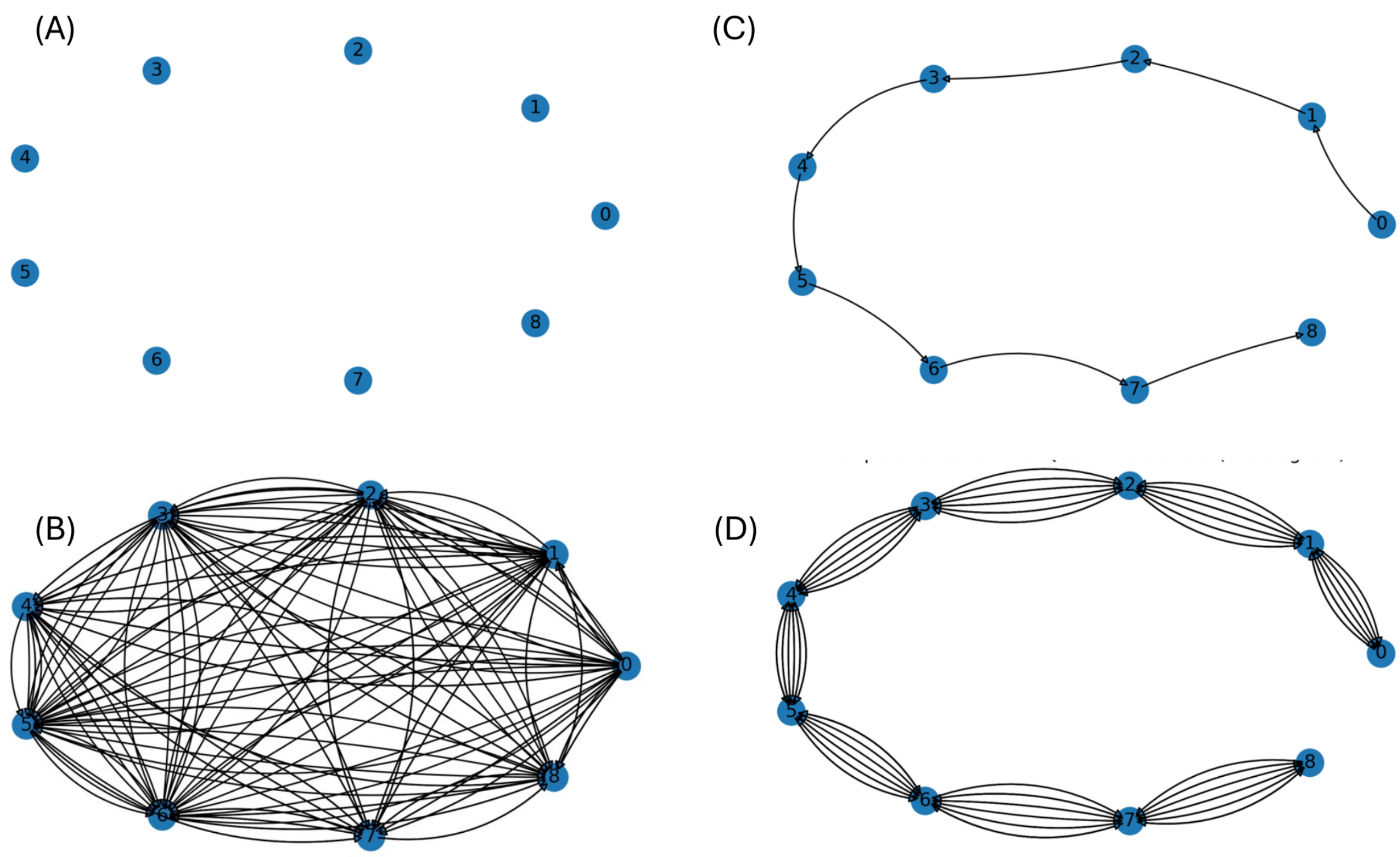

**Figure 9**: Feature-map two-qubit gates represented as directed graphs for the Z feature map without feature-map entanglement (A), high-entanglement all-to-all ZZ feature map (B), Z feature map with low-depth interleaved entanglement (C), and Z feature map with high-depth interleaved entanglement (D).

### *C.* Graph-theoretical measures of qubit interaction patterns

To characterize the tested two-qubit feature-map gate patterns, we computed standard graph-theoretic measures by representing directed two-qubit interactions between qubits as edges (Figure 9(A)-(D)). The graph diameter, average node degree, total number of directed interactions or steps, multiplicity, and reciprocity for each of these entanglement pattens is reported in Table 3. The graph diameter is the maximum shortest path connecting one qubit to another, while the average node degree is the mean number of interactions in which a qubit participates. Multiplicity measures redundancy: for example, two directed interactions from qubit *a* to qubit *b* give multiplicity 2 for the ordered pair (a,b). Reciprocity measures mutual interaction and is defined as the fraction of directed interactions that are bidirectional. Among the four configurations tested, the high-entanglement ZZ feature map has the highest unique connectivity, with 36 unique pair interactions and average degree 4.00. Its all-to-all topology gives graph diameter 1, so any qubit interacts directly with any other qubit at the feature-map level. The high-depth interleaved configuration has a comparable number of total directed steps but fewer unique qubit pairs, indicating that depth alone does not reproduce the same interaction topology as the all-to-all ZZ feature map.

**Table 3**. *Topological properties of entanglement patterns*

| Feature-map configuration | **Graph Diameter** | **Avg. Node Degree** | **Multiplicity** | Recip. | Directed steps (unique pairs) |
|---|---|---|---|---|---|
| Z feature map baseline (1 rep; no feature-map entanglement) | 0 | 0.00 | 0 | 0 | 0 |
| ZZ feature map high entanglement (2 reps; all-to-all) | 1 | 4.00 | 4 | 0 | 144 (36) |
| Z feature map + low-depth interleaved entanglement (1 pass) | 8 | 0.89 | 1 | 0 | 8 (8) |
| Z feature map + high-depth interleaved entanglement (9 passes) | 8 | 1.78 | 9 | 1 | 144 (16) |

These four configurations define the scope of the present graph summary. A systematic repetition-depth sweep, including one- and three-repetition ZZ maps, would be a natural follow-up and is not inferred from the present data.

## IV. Discussion

This study evaluates a sparse PRRS epitope-receptor binding classification task using a classical CNN benchmark and a fixed hybrid Embedding-QNN architecture in which only the feature-map entanglement configuration is varied. The main empirical finding is that unfiltered QNN training runs generally perform poorly, but validation-gated run filtering identifies subsets of runs whose test-set performance is comparable to the CNN benchmark. Within the four tested feature-map configurations, the all-to-all high-entanglement ZZ feature map appears to provide the most favorable balance between training-set fit and test-set performance: it maintains competitive test-set accuracy while producing a lower training AUAC and the strongest test/training AUAC ratio. This result supports the view that the topology of feature-map entangling gates, not only the total number of two-qubit gates, can influence performance in this particular sparse-data setting. The high-depth interleaved feature map contains a similar number of two-qubit gates but does not reproduce the same apparent reduction in overfit. At the same time, these simulations are noiseless statevector simulations. They do not model hardware decoherence, stochastic gate noise, or readout error, and therefore cannot by themselves support claims about beneficial physical noise on real NISQ devices. Prior work motivates future noise-aware and hardware-based tests [Preskill, 2018; Wang and Liu, 2024], but the present results should be interpreted as an analysis of feature-map design under deterministic simulation.

The results also call for a careful interpretation of generalization. Because the dataset contains only 80 labeled examples from a specific PRRS/SLA screening problem, we do not claim that QML generally outperforms classical learning or that entanglement generically improves generalization across tasks. Rather, the evidence is narrower: in this dataset and with this architecture, the validation-filtered high-entanglement ZZ feature map preserved test-set performance while fitting the training data less aggressively than the CNN and the other QNN feature-map configurations. This observation is compatible with theoretical work showing that QML generalization can be favorable in some small-data regimes [Caro et al., 2022], but it should also be read alongside results emphasizing that generalization depends on the data distribution, the number and type of trainable gates, and the structure of the learned task [Gil-Fuster et al., 2024]. Banchi et al. [2021] similarly frame the approximation-generalization trade-off in quantum-information terms, underscoring that finite-data limitations cannot be bypassed merely by choosing a quantum model. Thus, the present data support feature-map entanglement topology as a useful design variable for further study, not as a standalone explanation of broad generalization advantage.

## V. Conclusion

In summary, this work suggests that hybrid QML may be a useful exploratory tool for vaccine antigen screening given sparse data, provided its claims are framed at the level supported by this data. On the PRRS epitope-receptor dataset studied here, the fixed Embedding-QNN architecture did not perform well when all random initializations and data splits were averaged. However, validation-gated filtering identified QNN runs with test-set accuracy comparable to a simple CNN benchmark, and the high-entanglement all-to-all ZZ feature map produced the most favorable AUAC profile among the tested configurations. This approach is potentially valuable for screening of candidate epitopes because each additional high-fidelity docking or experimental measurement can be highly resource-intensive, and methods that extract usable signal from small labeled sets could help prioritize candidates quickly for subsequent higher-fidelity validation. The results also indicate that the structure of the feature map is impactful: two configurations with similar two-qubit-gate counts can behave differently when their interaction topologies differ. Future work should expand the dataset, test additional repetition depths and feature-map families, report uncertainty for AUAC and other aggregate metrics, include ROC-AUC and precision-recall analyses, evaluate calibration, and compare noiseless simulation against explicit noise models and hardware execution. Under those next tests, the promising behavior of the ZZ feature-map configuration can be assessed as part of a rigorous pipeline for triaging vaccine-antigen candidates under severe data constraints.

## VI. Data availability

The datasets used and/or analyzed during the current study are available from the corresponding author upon reasonable request.

## VII. References

Preskill J. Quantum Computing in the NISQ era and beyond. Quantum. 2018;2:79.

McClean JR, Boixo S, Smelyanskiy VN, Babbush R, Neven H. Barren plateaus in quantum neural network training landscapes. Nature Communications. 2018;9:4812.

Wang Y, Liu J. A comprehensive review of quantum machine learning: from NISQ to fault tolerance. Reports on Progress in Physics. 2024 Sep 25.

Caro MC, Huang HY, Cerezo M, Sharma K, Sornborger A, Cincio L, Coles PJ. Generalization in quantum machine learning from few training data. Nature Communications. 2022 Aug 22;13(1):4919.

Gil-Fuster E, Eisert J, Bravo-Prieto C. Understanding quantum machine learning also requires rethinking generalization. Nature Communications. 2024 Mar 13;15(1):2277.

Banchi L, Pereira J, Pirandola S. Generalization in quantum machine learning: A quantum information standpoint. PRX Quantum. 2021 Nov 1;2(4):040321.

von Lilienfeld OA, Müller KR, Tkatchenko A. Exploring chemical compound space with quantum-based machine learning. Nature Reviews Chemistry. 2020;4:347-358.

Haque A, Kumar V, Khan SN, Kim JJ. Quantum intelligence in drug discovery: Advancing insights with quantum machine learning. Drug Discovery Today. 2025 Sep 2:104463.

Shu R, Liu B, Xiong Z, Cui X, Li Y, Cui W, Yung MH, Qiao N. Quantum-inspired machine learning for molecular docking. arXiv preprint arXiv:2401.12999. 2024 Jan 22.

Jeong SG, Moon KH, Hwang WJ. Hybrid Quantum Neural Networks for Efficient Protein-Ligand Binding Affinity Prediction. arXiv preprint arXiv:2509.11046. 2025 Sep 14.

Pezeshkian W, Grünewald F, Narykov O, Lu S, Arkhipova V, Solodovnikov A, Wassenaar TA, Marrink SJ, Korkin D. Molecular architecture and dynamics of SARS-CoV-2 envelope by integrative modeling. Structure. 2023 Apr 6;31(4):492-503.

Zhang N, Qi J, Feng S, Gao F, Liu J, Pan X, Chen R, Li Q, Chen Z, Li X, Xia C. Crystal structure of swine major histocompatibility complex class I SLA-1*0401 and identification of 2009 pandemic swine-origin influenza A H1N1 virus cytotoxic T lymphocyte epitope peptide. Journal of Virology. 2011 Nov 15;85(22):11709-11724.

Lunney JK, Ho CS, Wysocki M, Smith DM. Molecular genetics of the swine major histocompatibility complex, the SLA complex. Developmental & Comparative Immunology. 2009 Mar 1;33(3):362-374.

Ho CS, Lunney JK, Ando A, Rogel-Gaillard C, Lee JH, Schook LB, Smith DM. Nomenclature for factors of the SLA system, update 2008. Tissue Antigens. 2009 Apr;73(4):307-315.

Abramson J, Adler J, Dunger J, Evans R, Green T, Pritzel A, Ronneberger O, Willmore L, Ballard AJ, Bambrick J, Bodenstein SW, et al. Accurate structure prediction of biomolecular interactions with AlphaFold 3. Nature. 2024 Jun 13;630(8016):493-500.

Jiang L, Zhang K, Zhu K, Wang Y, Kang Y, Hou T. Revisiting Protein-Protein Docking: A Systematic Evaluation Framework. Journal of Chemical Information and Modeling. 2025 Sep 18.

Alotaiq N, Dermawan D. Evaluation of Structure Prediction and Molecular Docking Tools for Therapeutic Peptides in Clinical Use and Trials Targeting Coronary Artery Disease. International Journal of Molecular Sciences. 2025 Jan 8;26(2):462.

Honorato RV, Trellet ME, Jiménez-García B, Schaarschmidt JJ, Giulini M, Reys V, Koukos PI, Rodrigues JP, Karaca E, van Zundert GC, et al. The HADDOCK2.4 web server for integrative modeling of biomolecular complexes. Nature Protocols. 2024 Nov;19(11):3219-3241.

Otsu N. A threshold selection method from gray-level histograms. IEEE Transactions on Systems, Man, and Cybernetics. 1979 Jan;9(1):62-66.

Kjeldsberg F, Munim ZH, Bustgaard M, Bhagat S, Lindroos E, Haavardtun P. Sensitivity of predictive performance assessment accuracy in varying k-fold cross validation. 2025;71.

Chen C, Isa NA, Liu X. A review of convolutional neural network based methods for medical image classification. Computers in Biology and Medicine. 2025 Feb 1;185:109507.

Zabin M, Choi HJ, Islam MM, Uddin J. Impact of Tuning Parameters in Deep Convolutional Neural Network Using a Crack Image Dataset. arXiv preprint arXiv:2506.03184. 2025 May 30.

Gallifant J, Chen S, Sasse K, Aerts H, Hartvigsen T, Bitterman DS. Sparse autoencoder features for classifications and transferability. arXiv preprint arXiv:2502.11367. 2025 Feb 17.

Singh N, Pokhrel SR. Modeling Feature Maps for Quantum Machine Learning. arXiv preprint arXiv:2501.08205. 2025 Jan 14.

Schuld M, Bocharov A, Svore KM, Wiebe N. Circuit-centric quantum classifiers. Physical Review A. 2020 Mar;101(3):032308.

Havlíček V, Córcoles AD, Temme K, Harrow AW, Kandala A, Chow JM, Gambetta JM. Supervised learning with quantum-enhanced feature spaces. Nature. 2019 Mar 14;567(7747):209-212.

Anand A. On the power of interleaved low-depth quantum and classical circuits. Doctoral dissertation, University of Waterloo; 2022.

Imambi S, Prakash KB, Kanagachidambaresan GR. PyTorch. In: Programming with TensorFlow: Solution for Edge Computing Applications. Cham: Springer International Publishing; 2021. p. 87-104.

Sahin ME, Altamura E, Wallis O, Wood SP, Dekusar A, Millar DA, Imamichi T, Matsuo A, Mensa S. Qiskit Machine Learning: an open-source library for quantum machine learning tasks at scale on quantum hardware and classical simulators. arXiv preprint arXiv:2505.17756. 2025 May 23.

Kingma DP, Ba J. Adam: A Method for Stochastic Optimization. International Conference on Learning Representations (ICLR); 2015.

Powell MJD. An efficient method for finding the minimum of a function of several variables without calculating derivatives. The Computer Journal. 1964 Jan 1;7(2):155-162.

Li S, Xia Y, Xu Z. Simultaneous perturbation stochastic approximation: towards one-measurement per iteration. Numerical Algorithms. 2023 Nov;94(3):1085-1101.

Brisebois AE, Broderick J, Khatooni Z, Wilson HL, Rayan S, Broderick G. Identifying Protein Co-regulatory Network Logic by Solving B-SAT Problems through Gate-based Quantum Computing. arXiv preprint arXiv:2504.09365. 2025 Apr 12.

Mandviwalla A, Ohshiro K, Ji B. Implementing Grover's algorithm on the IBM quantum computers. In: 2018 IEEE International Conference on Big Data (Big Data). IEEE; 2018 Dec 10. p. 2531-2537.

Abane A, Cubeddu M, Mai VS, Battou A. Entanglement routing in quantum networks: A comprehensive survey. IEEE Transactions on Quantum Engineering. 2025 Feb 11.

Park J. To Entanglement and Beyond: Explaining Superior Generalizability of Quantum Neural Networks. Proceedings of Quantum Techniques in Machine Learning (QTML2024), University of Melbourne, Melbourne, Australia; 2024 Nov 25-29.

**Acknowledgment**

This work was supported by the University of Saskatchewan's Centre for Quantum Topology and Its Applications (quanTA) and by the Vaccine and Infectious Disease Organization (VIDO). VIDO receives operational funding from the Canada Foundation for Innovation (CFI) through the Major Science Initiatives Fund and from the Government of Saskatchewan through Innovation Saskatchewan and the Ministry of Agriculture. The quanTA Centre's high-performance computational work has been advanced through a CFI John R. Evans Leaders Fund grant while access to quantum computing resources through IBM Quantum and PINQ2 has been made possible by a PrairiesCan Regional Innovation Ecosystems (RIE) contract (both awarded to SR). We give special thanks to members of IBM Quantum, in particular Nick Bronn, Alexandre Choquette, Julien Chosson, Genya Crossman, Paul Gulyas, and Sean Wagner, as well as to Marie-Eve Boulanger, Eric Capelle, Gael Humbert, Maxime Marchand, Marie-Claude Messier, and Mathieu Thery of PINQ2, for their tremendous help in facilitating our quantum hardware access and helping us to make the most of it scientifically. We thank the University of Saskatchewan Advanced Research Computing (ARC) team for their efforts and responsiveness in creating an excellent local environment for this computational work. This article is published with the permission of the Director of VIDO.

**Author information**

Authors and Affiliations listed above

**Contributions**

All authors contributed equally to this work.

**Corresponding authors**

Inquiries should be directed to the corresponding author(s).

**Ethics declarations**

Competing interests

The authors declare no competing interests.

**Additional information**

NA.